\documentclass[aps,pra,preprint,superscriptaddress,tightenlines]{revtex4}

\usepackage{graphicx} 
\usepackage{amsmath}
 
\newcommand{\beq}{\begin{equation}}
\newcommand{\enq}{\end{equation}}
\begin{document}

\title{Pairing in a three component Fermi gas}
\author{T. Paananen}
\email{Tomi.Paananen@helsinki.fi}
\affiliation{Department of Physical Sciences, University of Helsinki, 
PO Box 64, 00014 University of Helsinki,  Finland}
\affiliation{Nanoscience Center, Department of Physics,
PO Box 35, 40014 University of Jyv\"{a}skyl\"{a}, Finland}
\author{J.-P. Martikainen}
\affiliation{Department of Physical Sciences, University of Helsinki, 
PO Box 64, 00014 University of Helsinki,  Finland}
\author{P. T\"{o}rm\"{a}}
\affiliation{Nanoscience Center, Department of Physics,
PO Box 35, 40014 University of Jyv\"{a}skyl\"{a}, Finland}
\date{\today}
\begin{abstract}
We consider pairing in a three-component gas of degenerate fermions. In particular, we solve
the finite temperature mean-field theory of an interacting gas for a system where both interaction
strengths and fermion masses can be unequal. At zero temperature we find a
a possibility of a quantum phase transition between states associated with pairing 
between different pairs of fermions.
On the other hand, finite temperature behavior of the three-component system reveals 
some qualitative differences from the two-component gas: for a range of parameters
it is possible to have two different critical temperatures. The lower one corresponds
to a transition between different pairing channels, while the higher one corresponds
to the usual superfluid-normal transition. We discuss how these phase transitions could be observed in 
ultracold gases of fermionic atoms.
\end{abstract}
\maketitle

\section{Introduction}
\label{sec:intro}

Recently, the possibility of fermion pairing in a two-component
system with unequal Fermi surfaces has attracted 
a considerable amount of attention~\cite{Casalbuoni2004a}. 
In an electronic system non-matched Fermi surfaces could
be due to the magnetic field interacting
with electron spins and in color superconductivity, due to unequal quark masses.
The newly realized strongly interacting superfluid Fermi 
gases~\cite{Jochim2003b,Greiner2003a,Regal2004a,Zwierlein2004a,Bartenstein2004b,Kinast2004a,Chin2004a,Kinast2005a,Zwierlein2005a} 
offer a promising playground for the study of pairing and
superfluidity, also with imbalanced Fermi energies.

In ultra cold degenerate gases, different components are typically atoms
in different internal states of an atom, but different isotopes can also
be considered. In ultra cold gases the mismatch of the Fermi surfaces
can be due to having an unequal number of atoms in different 
states. After the first experimental studies using a mixture of 
two internal states of $^{6}{\rm Li}$ 
atoms~\cite{Zwierlein2006a,Partridge2006a}, 
the properties of such 
systems, in a harmonic trapping potential, have been extensively discussed in the recent 
literature~\cite{Bedaque2003a,Castorina2005a,Mizushima2005b,Yang2005a,Sheehy2006a,Kinnunen2006a,Pieri2005a,Yi2006a,Chevy2006a,DeSilva2006a,Haque2006a}.

In principle, many different types of fermions can be trapped
in a same trap. In order to reach long lifetimes, the atoms should have 
favorable collisional properties, but current knowledge 
of these collisional properties is limited. However, mixtures 
of either Bose
condensed or fermionic atoms in a variety of different internal states
have been experimentally 
demonstrated~\cite{Myatt1997a,Stenger1998a,Modugno2002a,Roati2002a,Schmaljohann2004a}, and experimental effort towards realizing 
three-component Fermi gases has been initiated. For this reason
we explore the novel possibilities in a three-component 
Fermi gas by generalizing the BCS (Bardeen-Cooper-Schrieffer)  
theory into three components. 

Related three-component Fermi system
has been previously studied by Honerkamp and Hofstetter~\cite{Honerkamp2004a,Honerkamp2004b}
and more recently by Bedaque and D'Incao~\cite{Bedaque2006a}.
The first reference employs a Hubbard lattice Hamiltonian and is focused
on the rather special case 
when all fermions have the same mass and interact with 
different fermions (as well as with fermions of the same type)
with a single interaction strength. 
The setting used in the latter reference includes the possibility 
of unequal interaction strengths between various
components. Bedaque and D'Incao draw general 
qualitative conclusions on the symmetries of the possible
zero temperature phases, but all fermion masses are assumed equal.

In this paper  we discuss pairing in a three-component system  both
at zero as well as non-zero temperatures. Furthermore, we allow for the possibility
that the third fermion type has a different mass from the other
two and since a priori there is little reason to expect
identical scattering properties between different fermion components,
we formulate our theory assuming
different interaction strengths between
the components. In ultracold gases, such a system
could be realized by trapping, for instance, $^6Li$ and $^{40}K$ using atoms in two internal 
states for one of theses and one internal state for the other
In order to restrict the parameter space somewhat, we make
one simplifying assumption in our Hamiltonian.
Generally one expects three different interactions in a three-component
system. One between the first and the second component, one between 
the second and the third component,
and one  between the third and the first component. In this paper we
assume that the interaction between the third and the first component
is weaker than the other two interactions and can 
be ignored next to the dominant contributions. This restriction is simply 
a matter  of convenience as it restricts parameter space to manageable
proportions, but does not affect the qualitative picture we find.

We find that, within the BCS theory, 
the possible pairing always gives rise to just
one order parameter of broken $U(1)$ symmetry. Whether the pairing takes
place between the first and the second component ($1-2$ channel) or
between the second and the third component ($2-3$ channel) 
depends on the strengths of the interactions, atomic masses,  as well as 
on the differences between the chemical potentials
of different components. The unpaired component constitutes a normal Fermi sea.
At zero temperature there is
a possibility of a quantum phase-transition from 
a phase with pairing in the $1-2$ channel into
a phase with pairing in the $2-3$ channel as chemical potential
differences are varied appropriately. This transition is of the first order. 
Furthermore, as the temperature increases it is possible
to have a second order transition from a paired phase in the 
$2-3$ channel into a paired phase  in the $1-2$ channel. At even higher temperature,
there is another transition from the paired state into the normal state.

This paper is organized as follows. In Sec.~\ref{sec:Normal}
we discuss the normal three-component Fermi gas 
and the possibility of phase-separation in such a system.
In Sec.~\ref{sec:FE} we present the BCS-style mean-field theory
and discuss the qualitative behavior of the associated 
grand potential at zero temperature. In this section we present 
the energy landscape
of the grand potential for several different scenarios.
In Sec.~\ref{sec:GE} we find the solutions to the gap-equations
that correspond to the global minimum of the grand potential
for fermions  where the third component has
a different mass from the other two components.
We end with some concluding remarks in Sec.~\ref{sec:Conclusions}.

\section{Normal three component Fermi gas}
\label{sec:Normal}
We study a homogeneous three component Fermi gas whose components are
either atoms in different internal states or atoms of different isotopes. 
For now we assume that an interaction between the first and the the third component 
is sufficiently small,
so that it is justified to focus on only the interactions between the first and the second component
and the second and the third component. The second quantized  Hamiltonian for this system is therefore
\beq
\label{eq:hamilton1}
\begin{split}
H&=\int d{\bf r} \,\left(\sum_{\sigma=1,2,3} {\hat \psi}_{\sigma}^{\dagger}({\bf r})
\left(-\frac{\hbar^2 \nabla^2}{2m_{\sigma}}-\mu_{\sigma}\right){\hat\psi}_{\sigma}({\bf r})\right)\\
&+\tilde g_{12}\int d{\bf r} \,{\hat\psi}_{1}^{\dagger}({\bf r})
{\hat\psi}_{2}^{\dagger}({\bf r}){\hat\psi}_{2}({\bf r}){\hat\psi}_{1}({\bf r})
+\tilde g_{23}\int d{\bf r} \,{\hat\psi}_{3}^{\dagger}({\bf r})
{\hat\psi}_{2}^{\dagger}({\bf r}){\hat\psi}_{2}({\bf r}){\hat\psi}_{3}({\bf r}),
\end{split}
\enq
where ${\hat\psi}_{\sigma}({\bf r})$ and ${\hat\psi}_{\sigma}^\dagger({\bf r})$ 
are the usual field-operators which annihilate and create particles in state $\sigma$.
In addition, $m_{\sigma}$ is the atomic mass, $\tilde g_{12}$ is the interaction strength between the first and
the second component, and $\tilde g_{23}$ is the interaction strength between the second 
and the third component. In terms of the scattering lengths $a_{ij}$ 
these interaction strengths are
given by $\tilde g_{ij}=2\pi\hbar^2 a_{ij}/\mu$, where $\mu=m_im_j/(m_i+m_j)$ 
is the reduced mass of the scattering atoms.

Let us first  discuss a mixture of three normal fermion components.
If the mixture of normal three-component gas phase-separates spatially, 
it is clear that there is no possibility of a superfluid
where all three components are mixed together and the 
problem is that of a two-component Fermi gas surrounded by the normal
third component. 
Since the temperatures of interests  are very low compared to the 
Fermi energies of various components 
we can consider the zero-temperature limit, in which the free-energy density
$f$ of the mixture of normal Fermi gases is given by
\beq
\label{eq:free1}
f=\frac{3\hbar^2}{10}(6\pi^2)^{2/3}(\frac{n_1^{5/3}}{m_1}+\frac{n_2^{5/3}}{m_2}+\frac{n_3^{5/3}}{m_3})+ \tilde g_{12}n_1n_2
+\tilde g_{23}n_2n_3,
\enq
where $n_i$ are the component densities.

If both interaction strengths $\tilde g_{12}$, $\tilde g_{23}$ are positive then all three components
separate. The reason for this is that 
then the positive definite free-energy Eq.~\eqref{eq:free1} is
minimized when the interaction terms give vanishing contributions. If one of
the interactions is positive and the other is negative,
the components with attractive  interaction
mix and  the third component separates. Therefore the case when all
three components can coexist spatially occurs only when
both interactions are attractive. 
If one of the interactions is attractive then 
formally Eq.~\eqref{eq:free1} has a global minimum at infinite density of
the interacting components.
This global minimum is not  physically relevant since 
there is an energetic barrier separating this solution from the physically
relevant regime of densities. In fact the height of this barrier increases
as the interactions become weaker. From now on we will focus on the most interesting
region where both interactions are fairly large and negative.

\section{Grand potential of the three component Fermi gas}
\label{sec:FE}
The BCS-theory involves approximating the Hamiltonian by  the  mean-field Hamiltonian density
(in the ${\bf k}$-space)
\beq
\begin{split}
H_{BCS}&=\frac{1}{V}\sum_{\bf k}\sum_{\sigma=1,2,3}\left(\frac{\hbar^2 k^2}{2m_\sigma}-\mu_{\sigma}\right){\hat 
\psi}^{{\dagger}}_{\sigma{\bf k}}
{\hat \psi}_{\sigma{\bf k}}+\Delta_{12} {\hat \psi}^{{\dagger}}_{1,\bf k} {\hat \psi}^{{\dagger}}_{2,\bf -k}
+ \Delta^*_{12} {\hat \psi}_{2,\bf -k} {\hat \psi}_{1,\bf k}+ 
\\
&+\Delta_{23} {\hat \psi}^{{\dagger}}_{3,\bf {k}} 
{\hat \psi}^{{\dagger}}_{2,\bf -k}+
\Delta_{23}^*{\hat \psi}_{2,\bf -k} {\hat \psi}_{3,{\bf k}}
-\frac{|\Delta_{12}|^2}{g_{12}}-\frac{|\Delta_{23}|^2}{g_{32}},
\end{split}
\enq
where 
\[\Delta_{\sigma\sigma'}=\frac{g_{\sigma\sigma'}}{V}\sum_{\bf k}\langle {\hat \psi}_{\sigma\bf k}
{\hat \psi}_{\sigma'\bf -k}\rangle\] are the order parameters to be determined self-consistently.
As a short hand notation we define the vector ${\bf \Delta}=\left(\Delta_{12},\Delta_{23}\right)$
characterizing the state of the system.

Since the mean-field Hamiltonian is of second order in the operators it can be
easily diagonalized with a canonical transformation. In this way we can 
calculate  the grand potential (or free energy) 
\beq
\label{eq:fe_1}
\Omega\left(\Delta_{12},\Delta_{23}\right)=-k_BT\log[Tr(\exp(-\beta H_{BCS}))],
\enq
where $k_B$ is the Boltzmann constant and $\beta=1/k_BT$. In the continuum limit
the  grand potential is ultraviolet divergent. This divergence is caused by
the unphysical short distance behavior of the contact interaction and is removed, in the
usual way, by subtracting the divergent contribution from the grand potential.
We have done the diagonalization of the mean field Hamiltonian which can be done analytically, 
but in the general  case formulas are inconveniently long and not very informative. In order to get
an overview of the expected system behavior it is rather more instructive 
to focus on the behavior of the free-energy.

We choose our units by assuming that the first and the second components are 
$^{6}{\rm Li}$ atoms and by
using the ideal gas of the second component as a benchmark.
The unit of energy is then $\epsilon_F=\hbar^2k_{F,2}^2/2m_2$,
where $k_{F,2}=(6\pi^2n_2)^{1/3}$ is the Fermi wave-vector. Furthermore, we use the unit 
of length $1/k_{F,2}$. Using these units we define the dimensionless coupling strengths as
$g_{12}=2k_{F,2}a_{12}/\pi$ and $g_{23}=(m_2+m_3)k_{F,2}a_{23}/m_2\pi$.

\subsection{Grand potential when all fermions have the same mass}
\begin{figure}
\begin{tabular}{ll}
\includegraphics[width=0.50\columnwidth]{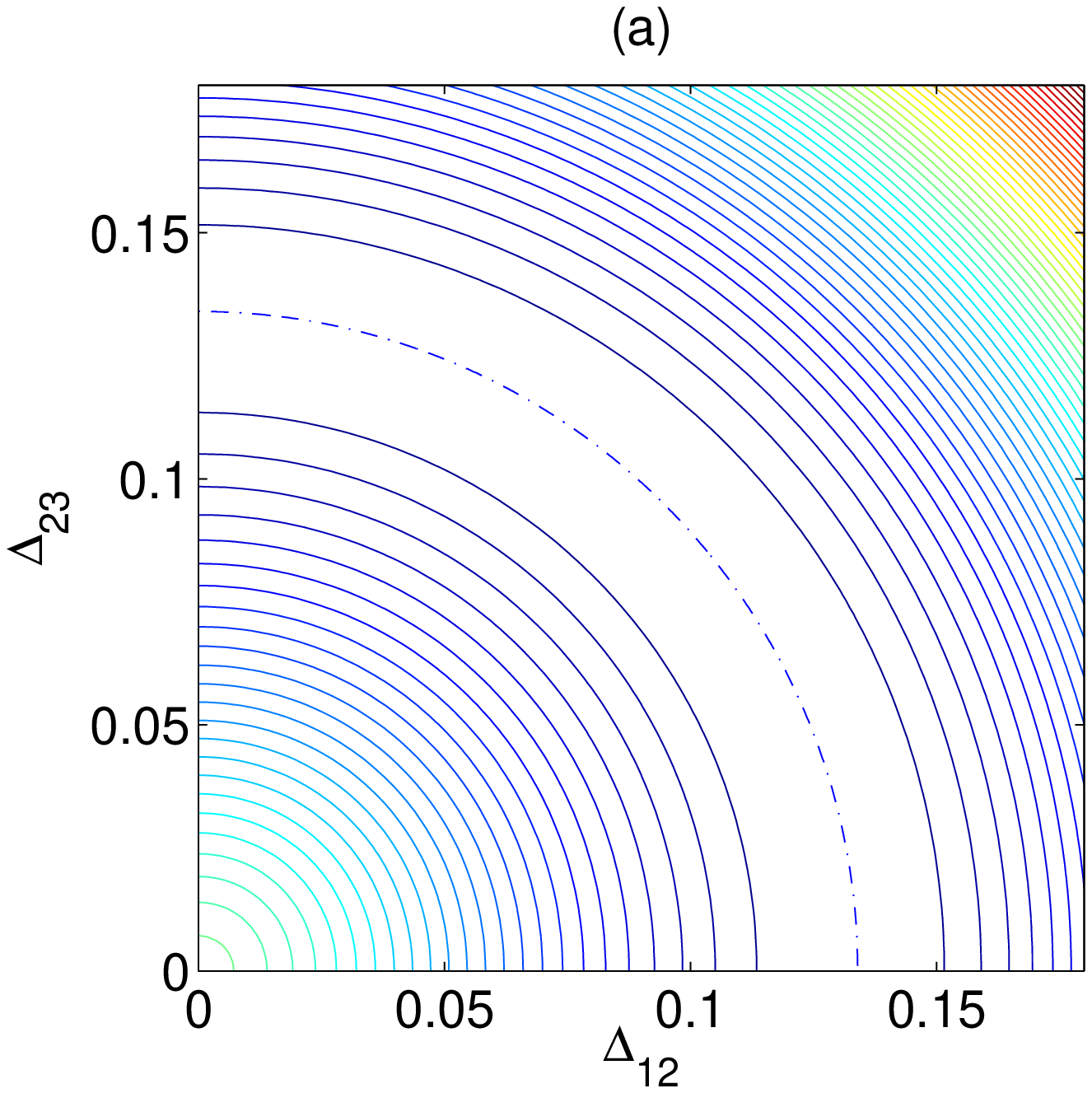} & \includegraphics[width=0.50\columnwidth]{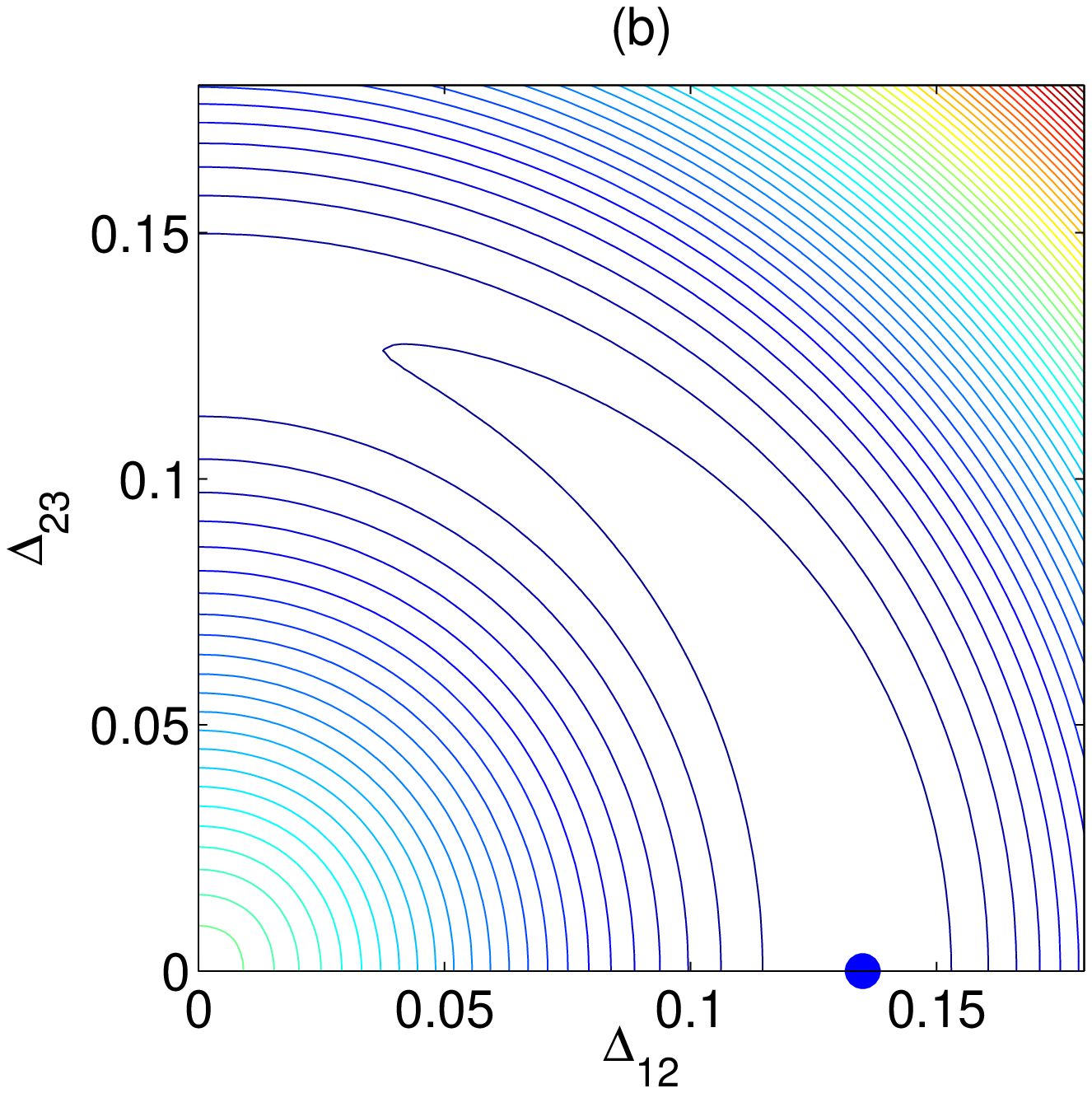}\\
\includegraphics[width=0.50\columnwidth]{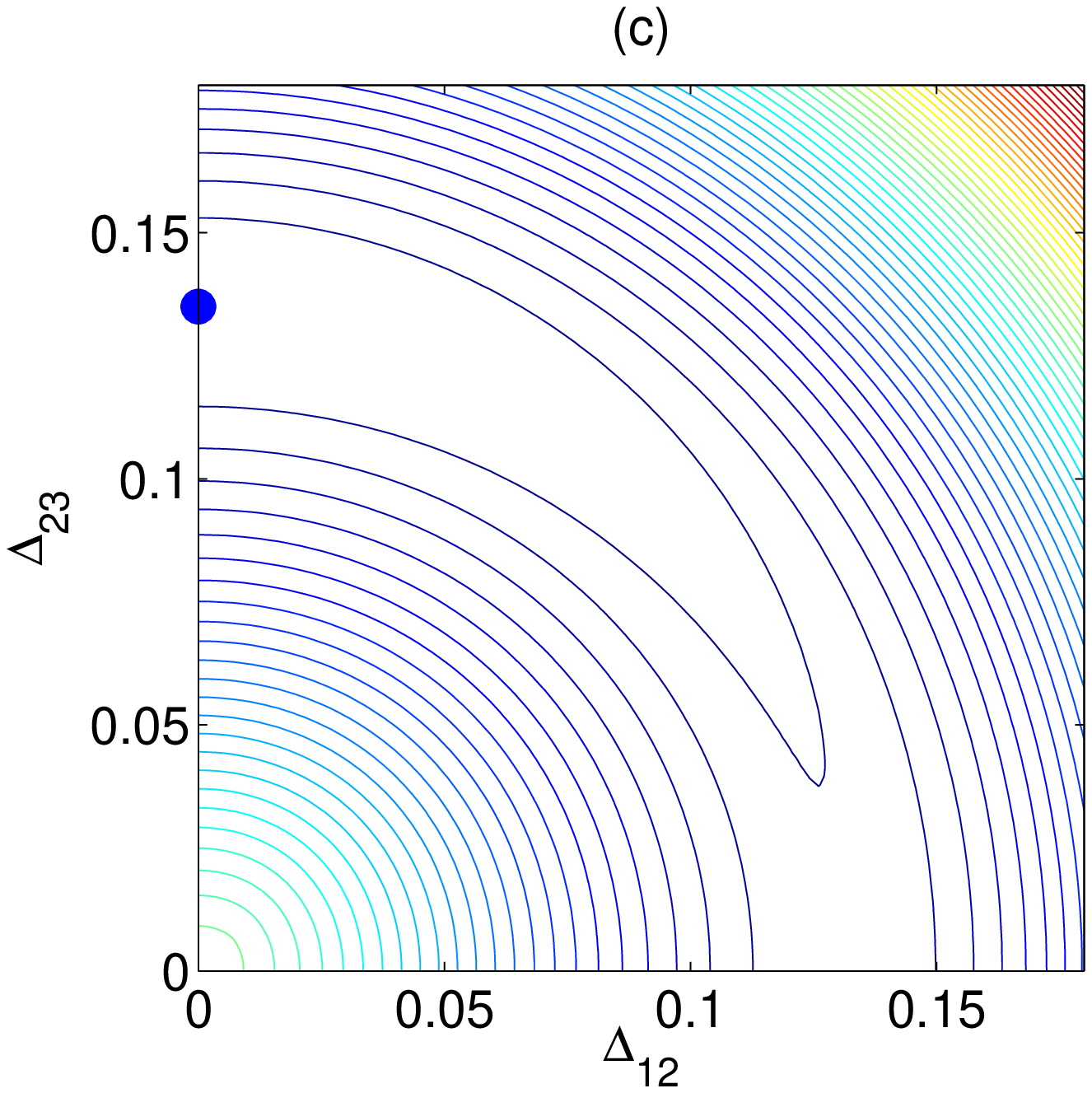} & \includegraphics[width=0.50\columnwidth]{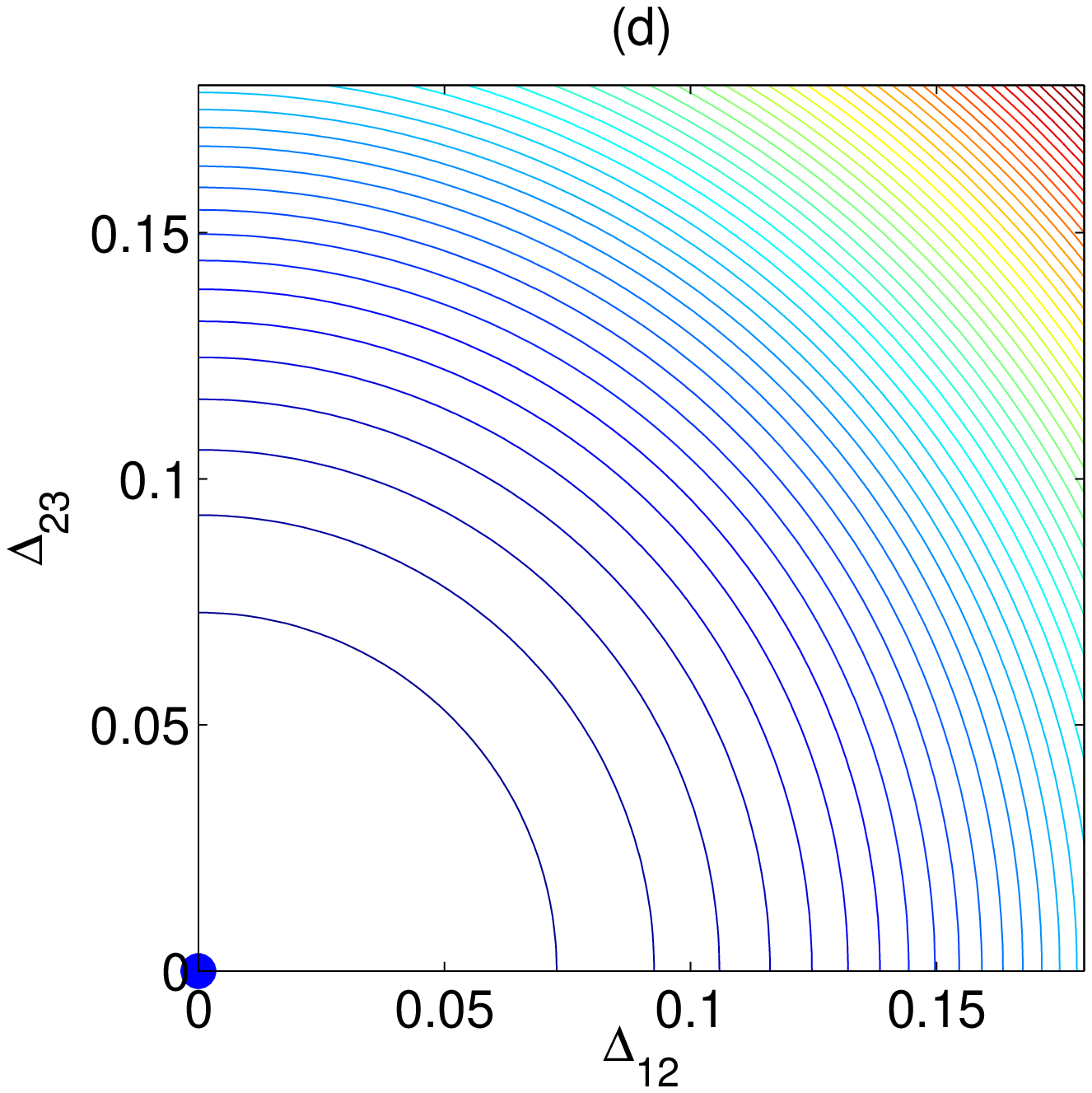}
\end{tabular}
\caption[Fig1]{(Color online) The free energy landscape as a function of
$\Delta_{12}$ and $\Delta_{23}$ when all three components have the same mass. 
The dash-dotted line in (a) and the filled circles in (b)-(d) show the location of the global minimum.
We used equal coupling strengths
$g_{12}=g_{23}=-0.50$ and
the figures (a)-(c) were calculated at zero temperature while
the figure (d) was calculated at $k_BT/\epsilon_F=0.08$ which is above the critical temperature. The chemical potentials
we such that
in the figure (a) $\mu_1=\mu_2=\mu_3=1$, in (b)  $\mu_1=1.01$, $\mu_2=1$, $\mu_3=0.99$,
in (c)  $\mu_1=0.99$, $\mu_2=1$,  $\mu_3=1.01$, and
finally in (d)  $\mu_1=\mu_2=\mu_3=1$.
}
\label{fig:fe_1}
\end{figure}

The Fig.~\ref{fig:fe_1} demonstrates the behavior of the free energy landscape as a function
of the gaps $\Delta_{12}$ and $\Delta_{23}$ when all fermion components have the same mass.
It is clear that when all chemical potentials are equal the free energy landscape
is only a function of $\Delta_{12}^2+\Delta_{23}^2$ and at zero temperature
displays a clear minimum corresponding to the solution of the BCS gap equation. 
With the parameters used in the figure, the minimum is located at
$\sqrt{\Delta_{12}^2+\Delta_{23}^2}\approx 0.135$, see the quarter sphere in Fig.~\ref{fig:fe_1}(a).

Any difference in the chemical potentials
(or coupling strengths) breaks the above symmetry and makes it energetically favorable
to have another one of the  phases ${\bf \Delta}=(\Delta_{12},0)$ or
${\bf \Delta}=(0,\Delta_{23})$. When interactions are the same, 
which of these alternatives is chosen depends on the 
average chemical potential between the paired components. The channel corresponding
to the higher average chemical potential has a lower free energy and 
is therefore physically realized. At zero temperature, the densities of the paired
components are equal, as is to be expected from the BCS-state.
At higher temperatures the normal state
${\bf \Delta}=(0,0)$ eventually becomes the free energy minimum.
The Figure~\ref{fig:fe_1}  also demonstrates 
that generally the grand potential has several possible stationary points.

\subsection{Grand potential when the fermion masses differ}
\begin{figure}
\begin{tabular}{ll}
\includegraphics[width=0.50\columnwidth]{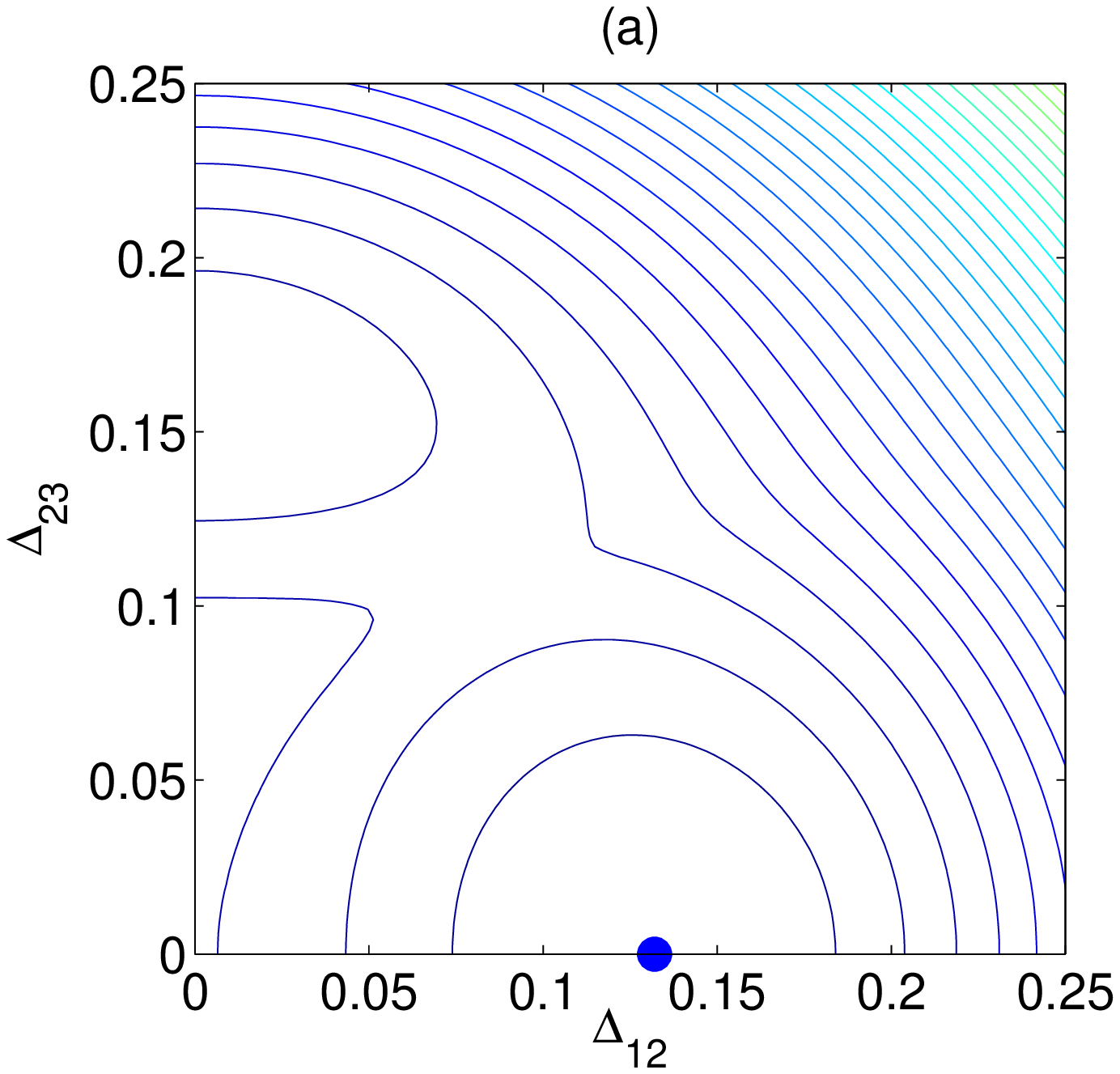} & \includegraphics[width=0.50\columnwidth]{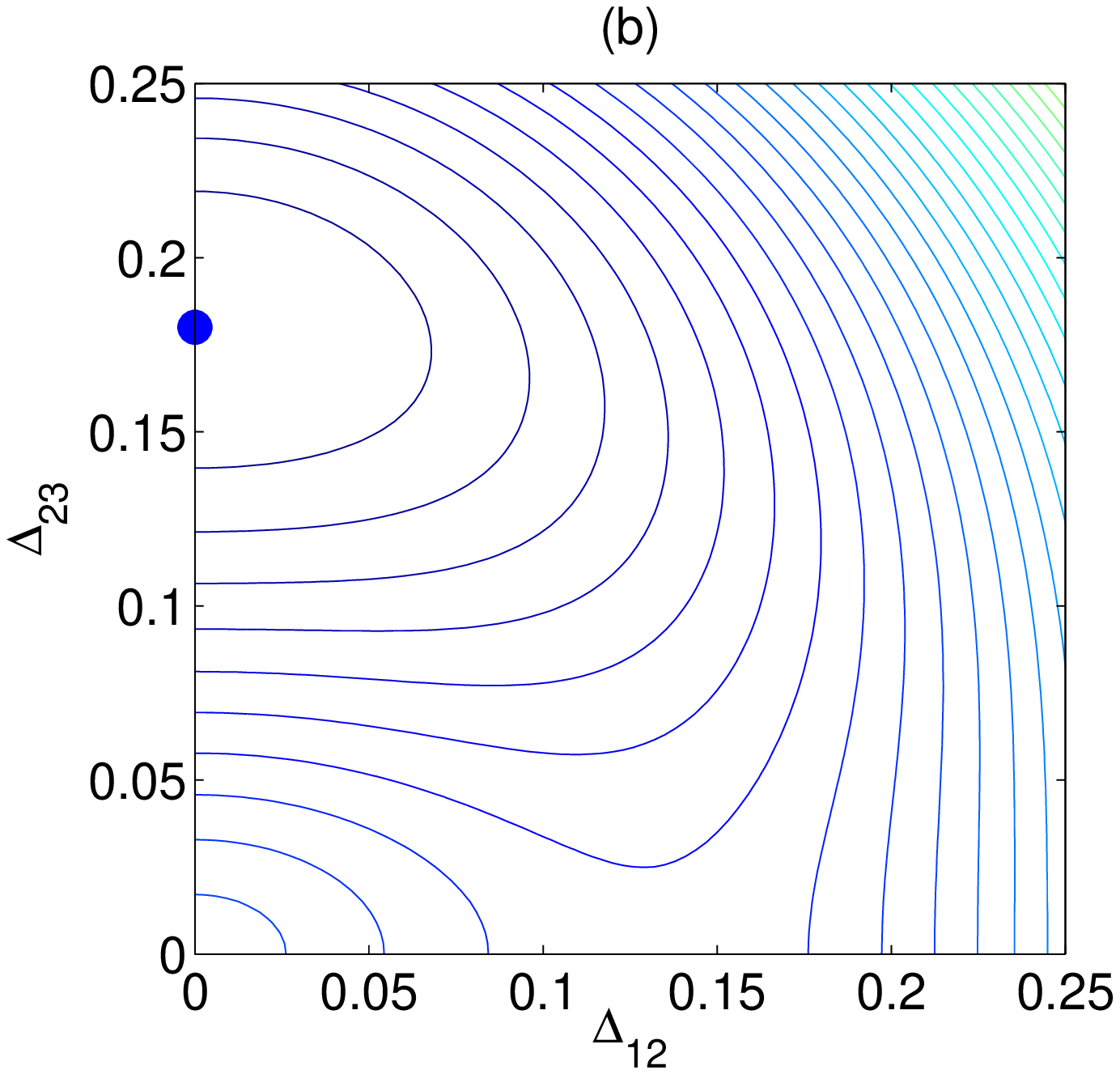}\\
\includegraphics[width=0.50\columnwidth]{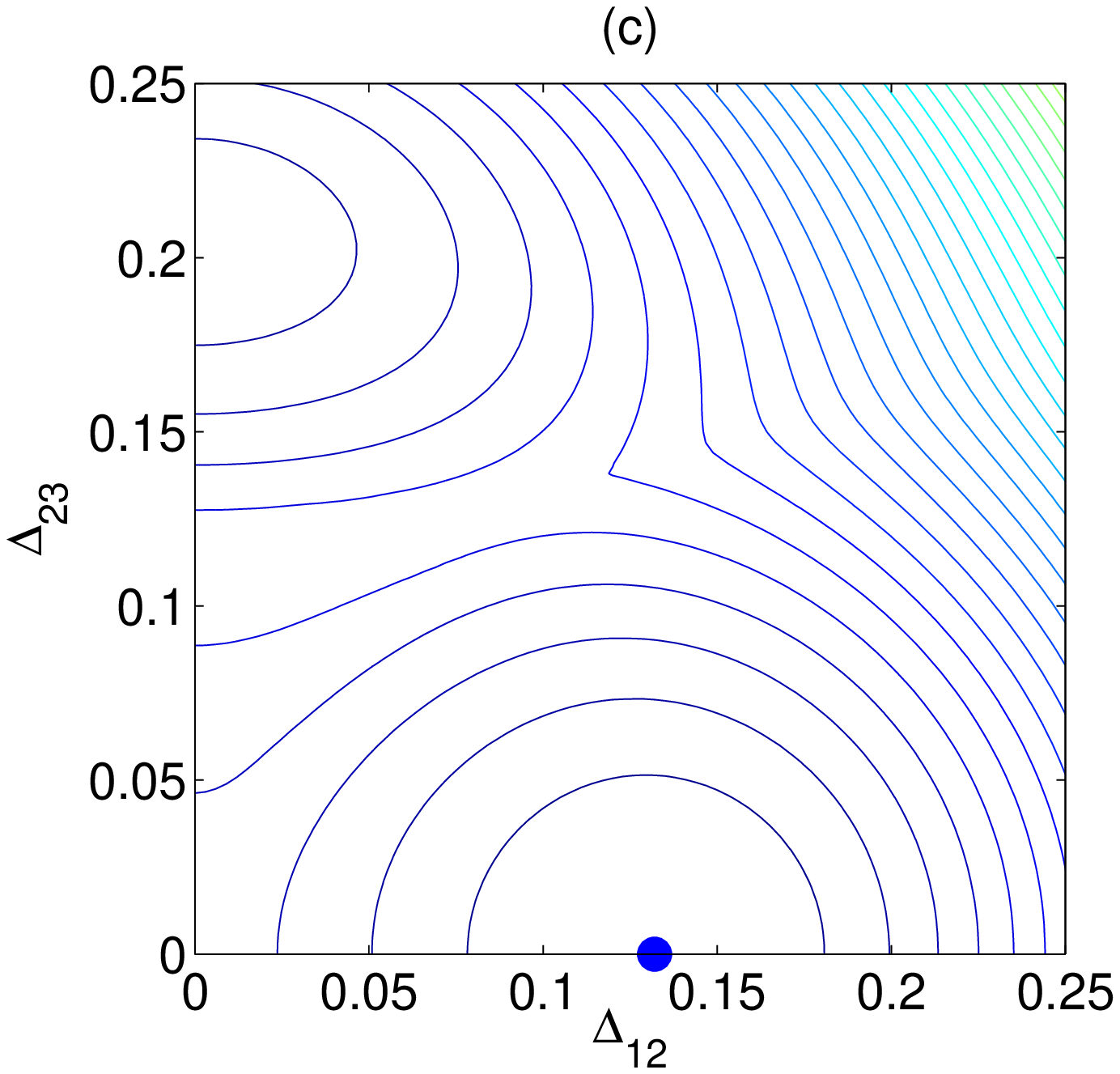} & \includegraphics[width=0.50\columnwidth]{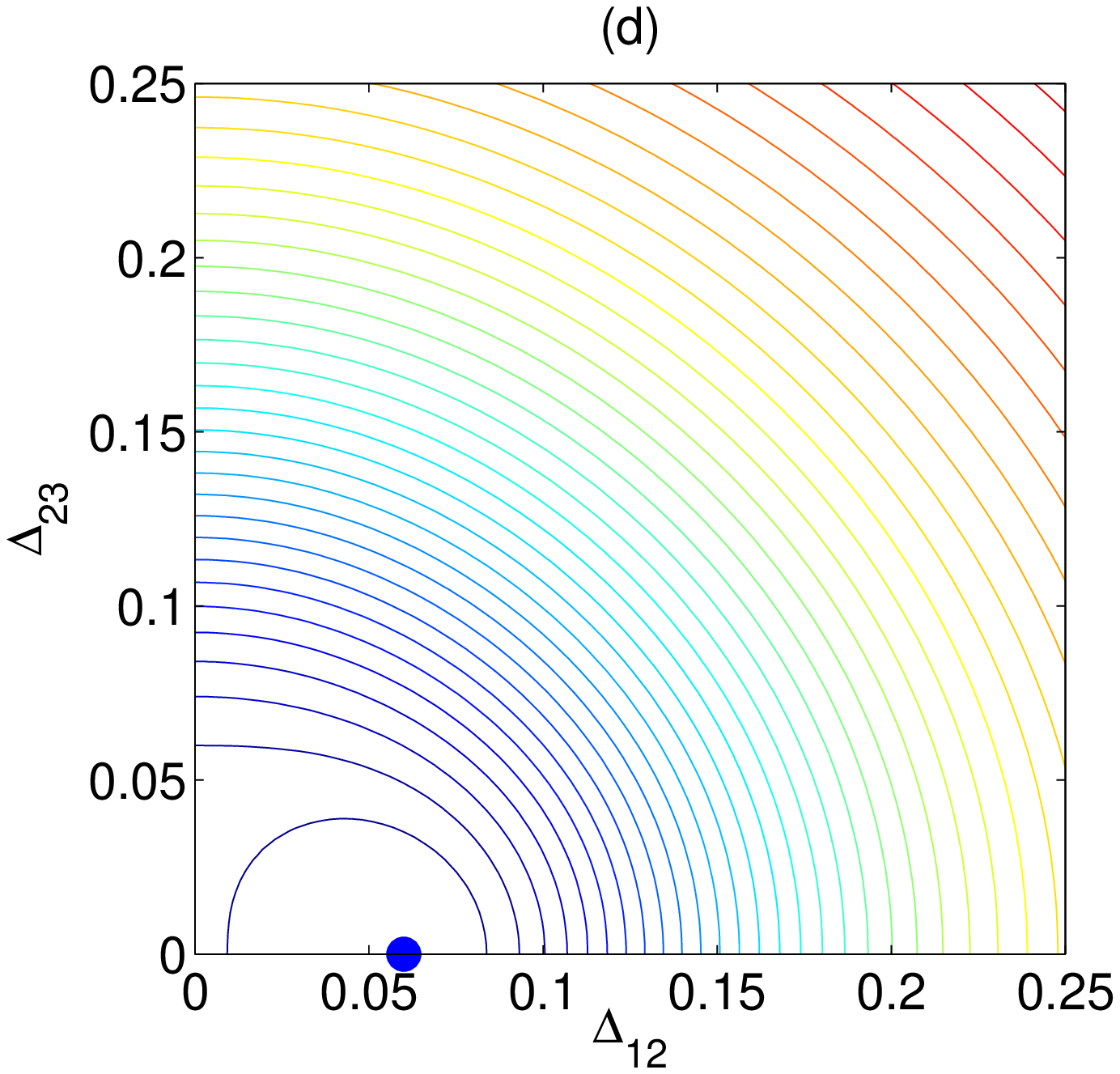}
\end{tabular}
\caption[Fig2]{(Color online)  The free energy landscape as a function of
$\Delta_{12}$ and $\Delta_{23}$ with the mass ratio $m_1/m_3=0.15$
and  $\mu_1=\mu_2=1.0$.
The filled circle shows a location of the global minimum.
The coupling strengths were $g_{12}=g_{23}=-0.5$ and
the figures (a)-(c) were calculated at zero temperature while
the figure (d) was calculated at $k_BT/\epsilon_F=0.07$.
In figure (a) $\mu_3=0.07$, in (b) $\mu_3=0.15$, in (c) $\mu_3=0.25$, and
in (d) $\mu_3=0.15$. (In the figure (d) we used a logarithmic scale
to calculate the contour plot in order to enhance the relevant features.)
The results show two types of transitions: a zero temperature quantum phase transition between
the two pairing channels (from (a) to (b), and from (b) to (c)), and a finite temperature 
second order transition between the pairing channels (from (b) to (d)).
}
\label{fig:fe_2}
\end{figure}

The different components could in principle have also different masses.
For concreteness we assume that the first and the second component are different internal states of
$^{6}{\rm Li}$ atoms while the third component is a
$^{40}{\rm K}$ atom. This therefore corresponds to the mass ratio $m_r=m_1/m_3=0.15$.
In the Fig.~\ref{fig:fe_2} we show some typical free energy landscapes
in this case, when the chemical potential of the third component is varied.
It is clear that the grand potential again has many different stationary points.

For small values of $\mu_3$ the pairing is only possible in the $1-2$ channel.
However, the Fermi surfaces are matched when $\mu_3=m_r$ and this is reflected
as a possibility of pairing in the $2-3$ channel around this value. 
The reason why
pairing in this channel is favored for matched Fermi surfaces even
when coupling strengths are equal, is due to the higher density of states
for atoms of higher mass. This higher density of states translates into
reduction in energy. Naturally if $g_{23}$ is reduced sufficiently
we enter a parameter region where pairing in the $2-3$ channel will
never take place. For the parameters used in the figure 
this happens when $|g_{23}|<0.33$.

Again we find that at zero temperature the densities of the paired
components are always equal.
It is also interesting to observe that by increasing the temperature, one
can induce a transition from the ${\bf \Delta}=(0,\Delta_{23})$ phase
into the ${\bf \Delta}=(\Delta_{12},0)$ phase before entering a normal state.

In summary, two transitions are visible: a zero temperature quantum phase transition between
the two pairing channels (transition from Fig.~\ref{fig:fe_2}(a) to Fig.~\ref{fig:fe_2}(b), and from Fig.~\ref{fig:fe_2}(b) to 
Fig.~\ref{fig:fe_2}(c)), and a finite temperature 
second order transition between the pairing channels (transition from Fig.~\ref{fig:fe_2}(b) to Fig.~\ref{fig:fe_2}(d)). 

\section{Solutions of the gap-equations of the three component Fermi gas}
\label{sec:GE}
For a given set of chemical potentials the self-consistent BCS-solution
is found at the minimum of the grand potential. At the extremum value of
the grand potential, the gap-equations
\beq
\frac{\partial \Omega\left(\Delta_{12},\Delta_{23}\right)}{\partial \Delta_{12}}=0
\enq
and 
\beq
\frac{\partial \Omega\left(\Delta_{12},\Delta_{23}\right)}{\partial \Delta_{23}}=0
\enq
are satisfied. These equations are satisfied at all extremal points of the grand potential,
including local minima, local maxima, and saddle points.
As we found out in the earlier section, the grand potential can have several extremal
points so care must be taken in ensuring that numerics converges to the physically
relevant global minimum of the grand potential rather than to a local stationary point.

\begin{figure}
\begin{tabular}{lll}
\includegraphics[width=0.333\columnwidth]{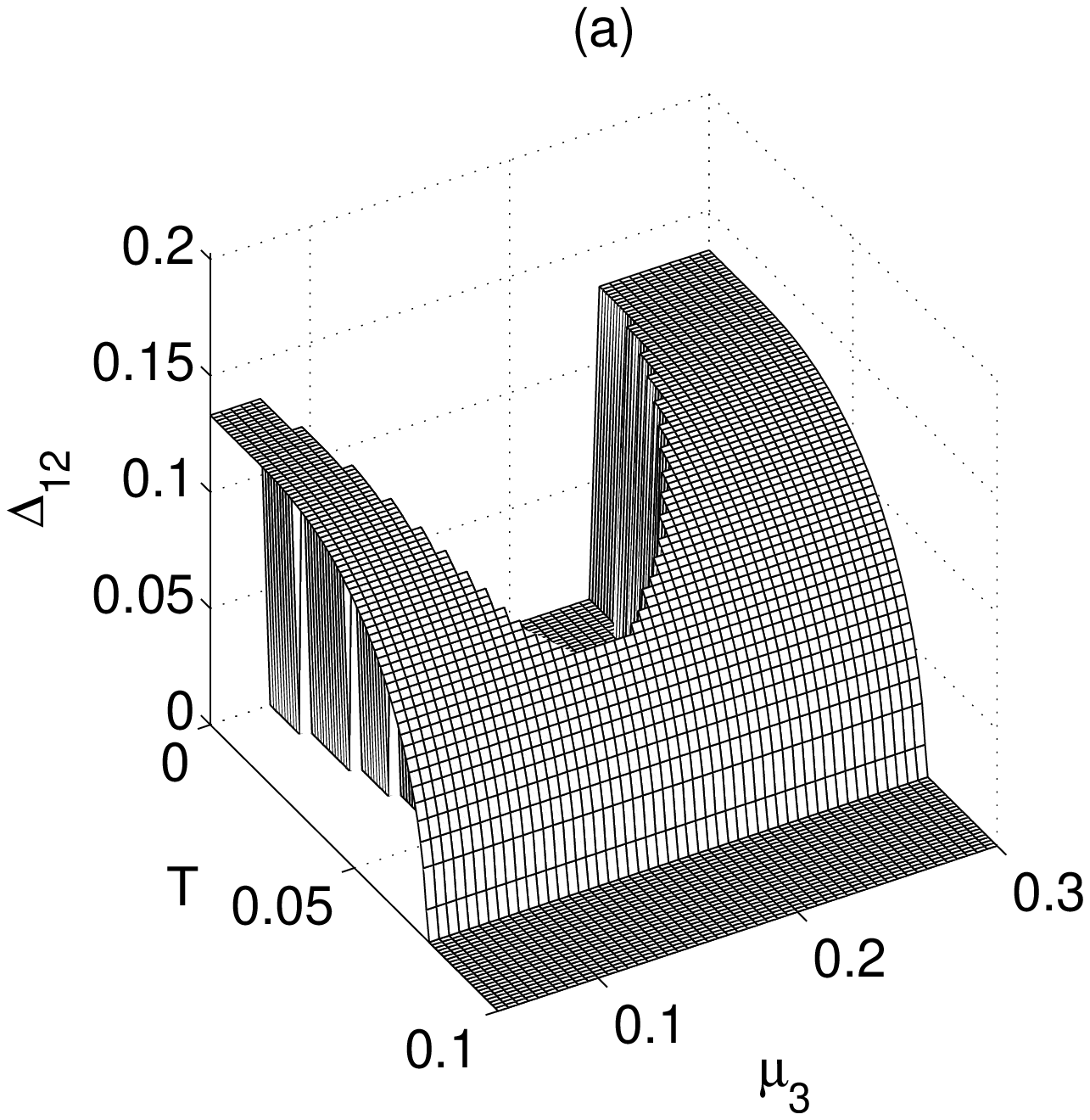} & \includegraphics[width=0.333\columnwidth]{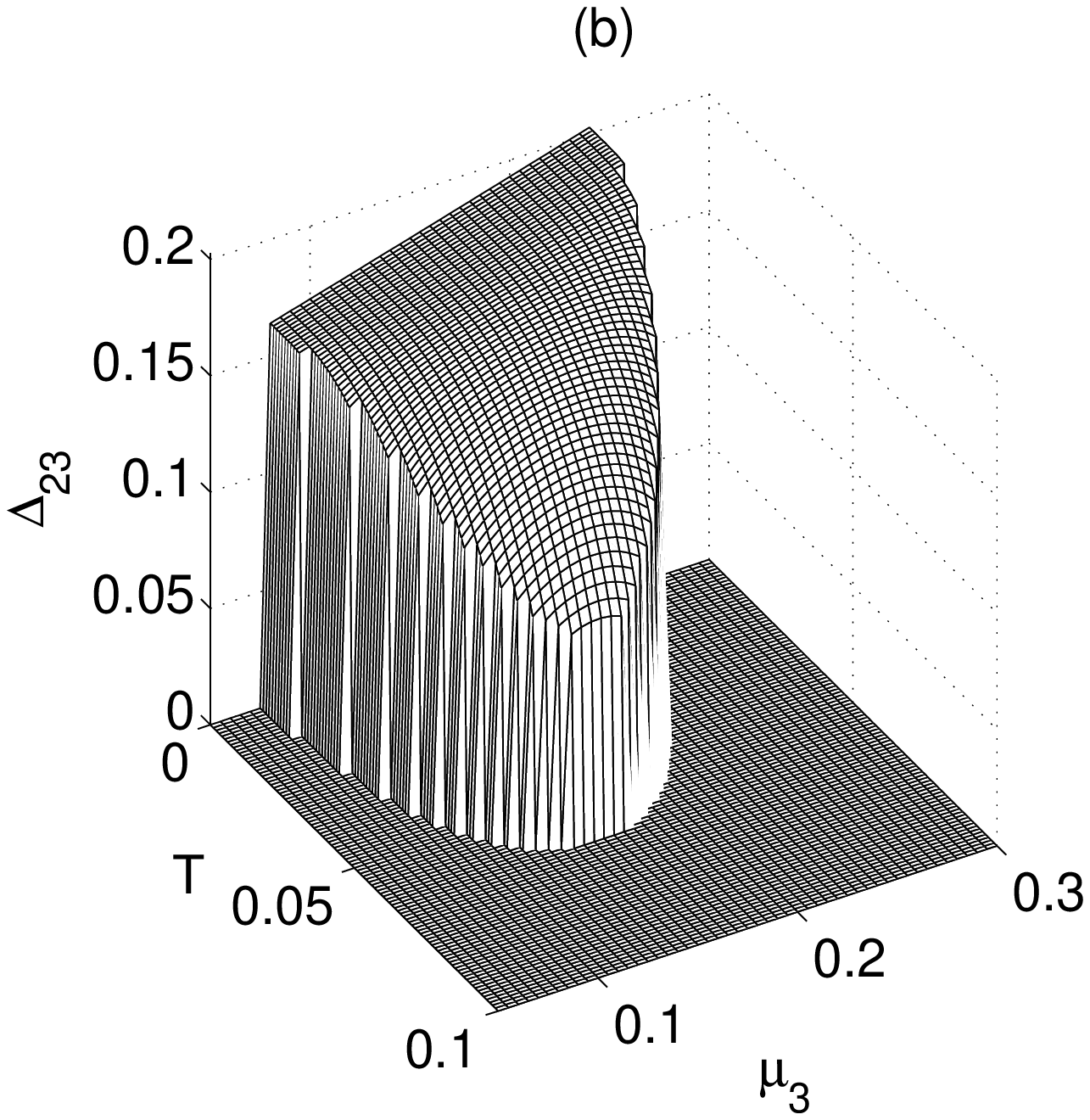}
& \includegraphics[width=0.333\columnwidth]{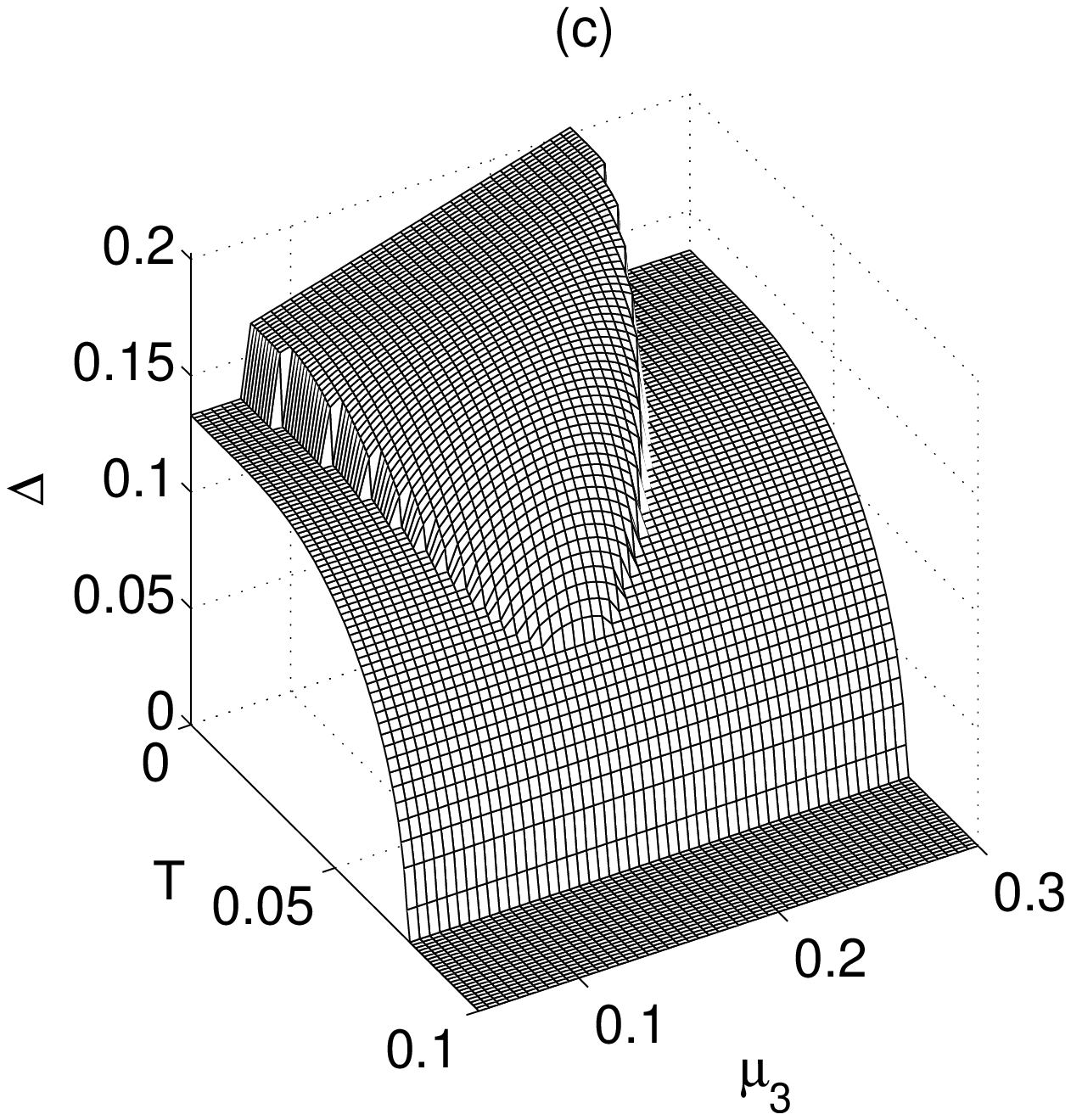}
\end{tabular}
\caption[Fig3]{The gap parameters as a function of temperature $T$ and of the third component
chemical potential $\mu_3$. We used the mass ratio $m_1/m_3=0.15$, $\mu_1=\mu_2=1$,
and the coupling strengths were $g_{12}=g_{23}=-0.5$. The figure (c) shows 
$\Delta=\sqrt{\Delta_{12}^2+\Delta_{23}^2}$ and
demonstrates clearly the sharp change in 
the order parameter at zero temperature (the quantum phase transition) as well as the smoother transition from the
$\Delta_{23}$ superfluid to $\Delta_{12}$ superfluid at a finite temperature.  
}
\label{fig:gaps}
\end{figure}

We have found the global minima of the free energy by solving the gap equations
under many different circumstances. Fig.~\ref{fig:gaps} shows typical results
for the gaps as a function of temperature and $\mu_3$. At zero temperature 
there exists a first order quantum phase transition, as the chemical potential
is varied, from the ${\bf\Delta}=(\Delta_{12},0)$ phase into the ${\bf\Delta}=(0,\Delta_{23})$
phase. We also find that with increasing temperature there can
be a second order transition from the ${\bf\Delta}=(0,\Delta_{23})$
phase into the ${\bf\Delta}=(\Delta_{12},0)$ phase, which then makes a transition
into the normal state at a higher temperature. However, whether this sequence of transitions
takes place or not, does depend on the strength of the interactions.
For weaker interactions it is possible that the ${\bf\Delta}=(0,\Delta_{23})$ phase is
not completely surrounded by the ${\bf\Delta}=(\Delta_{12},0)$  phase as it is in
Fig.~\ref{fig:gaps}. 

This overall behaviour can be understood in the following way: The pairing between 
atoms of different masses happens around the point where their Fermi momenta are nearly equal (following from
the condition $\mu_3/\mu_1 ~ m_1/m_3$). This pairing is favoured over the pairing with same mass components due to the higher
density of states for the more massive component. At finite temperature, the Fermi surfaces of both pairing components will 
be smeared out due to temperature - however, for components with different masses, this smoothing of the Fermi edge 
will happen differently (the higher mass component Fermi edge spreads more). Therefore, with increasing temperature, 
it becomes increasingly difficult to find a pairing partner with the same magnitude (but opposite sign) momentum, and also 
the advantage given by the higher density of states of the more massive component becomes less significant. This makes 
the pairing between the same mass components potentially more favourable at high temperatures.    

\section{Summary and conclusions}
\label{sec:Conclusions}
In this paper we have considered the mean-field theory 
of an interacting three-component Fermi gas. We found
that in the highly symmetric case 
when all fermion masses as well as interactions strengths are equal the
free energy is only a function of $\Delta_{12}^2+\Delta_{23}^2$.
However, any deviation from the most symmetric situation, whether it is
due to different atomic masses, different interaction strengths, or
different chemical potentials, breaks the above symmetry of 
the free energy and makes one of the paired states $(\Delta_{12}\neq0 ,\Delta_{23}=0)$,
$(\Delta_{12}=0 ,\Delta_{23}\neq 0)$ or the normal-state
$(\Delta_{12}=0 ,\Delta_{23}= 0)$
energetically favored.  Magnitudes of non-zero gaps
as well as the point where the transition to the normal-state occurs 
depend on parameters, but we have solved the global minima of the
free energy 
for fairly representative range of parameters.

When using a grand canonical ensemble the average densities are quantities
derived from the grand potential through $n_i=-\partial\Omega/\partial\mu_i$.
In this work we have explored the system behavior as a function of the chemical potentials
by minimizing the grand potential. It is therefore useful
to elaborate what kind of density ratios  different sets of chemical potentials
actually correspond to. For the case investigated in the earlier section with a mass
ratio $m_1/m_3=0.15$ at zero temperature, we find at 
$\mu_3=0.05$ pairing in the $1-2$ channel and density
ratios of $n_3/n_1=n_3/n_2\approx 0.26$. As the third component chemical potential
is increased to $\mu_3=0.15$, we find pairing in the $2-3$ channel
with density ratios $n_2/n_1=n_3/n_1\approx 0.72$. Finally, when 
$\mu_3=0.25$, the pairing is reverted back  to the $1-2$ channel and the mass ratios
are $n_3/n_1=n_3/n_2\approx 2.9$. Therefore, this transition could be realized simply by 
varying the density of one component and keeping the other two fixed. Note that transitions
could also be induced by changing the interaction strengths, i.e. the scattering lenghts, 
between the three components as discussed in~\cite{Bedaque2006a}. However, at the present the method used for
tuning the scattering lengths, i.e.\ the use of Feshbach resonances, does not allow to tune the different
scattering lengths independently whereas the densities can be varied completely independent of each other.
The change in the order gap related to the transitions
could be observed using RF-spectroscopy of the pairing 
gap~\cite{Chin2004a,Kinnunen2004b,Torma2000a} 
and corresponding condensate fraction 
by the pair projection method~\cite{Regal2004a,Zwierlein2004a}, 
and superfluidity could be directly demonstrated by the 
creation of
vortices~\cite{Zwierlein2005a} - all these can be studied selectively 
for any choice for the pairing between the three 
components.

In this paper we have for convenience limited ourselves
to the special case where interaction between the first
and the third component is small enough to be neglected.
This requirement can be relaxed easily and one 
could solve the most general case with unequal masses as well
as with unequal interactions. We have checked that
the presence of the omitted interaction  
between the first and the third component does not
change the qualitative picture presented in this paper.
For some parameter values the new gap-function
$\Delta_{13}$ can become non-zero, but otherwise the qualitative 
physics remains similar to the cases discussed in this paper.

Three component Fermi system has been studied theoretical
\cite{Honerkamp2004a,Honerkamp2004b,Bedaque2006a,Paananen2006a}.
The first reference employs a Hubbard lattice Hamiltonian and is focused
on the rather special case 
when all fermions have the same mass and interact with 
different fermions (as well as with fermions of the same type)
with a single interaction strength. 
The setting used in the latter reference includes the possibility 
of unequal interaction strengths between various
components. Bedaque and D'Incao draw general 
qualitative conclusions on the symmetries of the possible
zero temperature phases, but all fermion masses are assumed equal.

\begin{acknowledgments}
The authors thank T. Koponen and J. Kinnunen for useful discussions.
This work was supported by Academy of Finland and EUROHORCs 
(EURYI award, Academy project numbers 207083, 106299, and 205470).

\end{acknowledgments}

\bibliographystyle{apsrev}
\bibliography{bibli}

\end{document}